\documentclass[12pt]{iopart}
\usepackage{natbib} 
\bibpunct{[}{]}{;}{n}{,}{,}
\expandafter\let\csname equation*\endcsname\relax
\expandafter\let\csname endequation*\endcsname\relax
\usepackage{amsmath}
\usepackage{amssymb}
\usepackage{color}
\usepackage{graphicx}
\newcommand{\newblock}{\hskip .11em\@plus.33em\@minus.07em}
\begin{document}
\title
[Is the Universe anisotropic right now?]{Is the Universe anisotropic right now? Comparing the real Universe with the Kasner's space-time}
          
\author{S L Parnovsky$^{1,2}$}

\address{$^1$Taras Shevchenko National University of Kyiv, Astronomical observatory, Observatorna str. 3, Kyiv 04053, Ukraine}
\address{$^2$D\'epartement de Physique Th\'eorique, Universit\'e de Gen\`eve,
 24 quai Ernest-Ansermet,  CH-1211 Gen\`eve 4, Switzerland}
\ead{parnovsky@knu.ua}
\vspace{10pt}
\begin{abstract}
We investigate possible astronomical manifestations of space-time anisotropy. The homogeneous vacuum Kasner solution was chosen as a reference anisotropic cosmological model because there are no effects caused by inhomogeneity in this simple model with a constant degree of anisotropy. This anisotropy cannot become weak. The study of its geodesic structure made it possible to clarify the properties of this space-time. It showed that the degree of manifestation of anisotropy varies significantly depending on the travel time of the light from the observed object. For nearby objects, for which it does not exceed half the age of the universe, the manifestations of anisotropy are very small. Distant objects show more pronounced manifestations, for example, in the distribution of objects over the sky and over photometric distances. These effects for each of the individual objects decrease with time, but in general, the manifestations of anisotropy in the Kasner space-time remain constant due to the fact that new sources emerging from beyond the cosmological horizon.We analyse observable signatures of the Kasner-type anisotropy and compare it to observations. These effects were not found in astronomical observations, including the study of the CMB. We can assume that the Universe has always been isotropic or almost isotropic since the recombination era. This does not exclude the possibility of its significant anisotropy at the moment of the Big Bang followed by rapid isotropization during the inflationary epoch.
\end{abstract}

%
\vspace{2pc}

\submitto{\CQG}%

\section{Introduction}
This work has two main goals. The first is to study the properties of the Kasner's space-time for a better understanding of its physical meaning. The second is the application of the obtained results to the study of the question of whether the Universe can be anisotropic. Let us explain in more detail why we are studying these problems.

Many exact solutions starting from the cosmological singularity have been obtained within the framework of general relativity. Many of them are given in the book \citep{p1}. But very few of them can be used to describe the properties of the real Universe that we know from centuries of astronomical observation and in which we all exist, including the author and readers of this article.

Hardly anyone believes that, for example, the G\"odel universe (see \citep{p2}) is suitable for describing the world around us. When it comes to the real Universe, the circle of cosmological models sharply narrows. And the most adequate model for describing the space-time filled with real galaxies, quasars, CMB and other radiation is still the mainstream $\Lambda$CDM model. But this name usually means something more than the universe filled with matter, including cold dark matter (CDM), and dark energy ($\Lambda$) in one form or another. The vast majority of scientists implicitly add the requirement of isotropy and homogeneity of the model. So it should be called by its full name Friedmann-Lema\^{i}tre-Robertson-Walker (FLRW) $\Lambda$CDM model.

There are no problems in building an anisotropic homogeneous $\Lambda$CDM model. Such a model was obtained in \citep{p3} for the Bianchi type I space. The universe is currently almost isotropic in this model. The presence of matter or dark energy in any combination leads to the isotropization. It is especially significantly isotropized during the inflationary epoch. After its completion, the initial anisotropy, which is so significant at the anisotropic Big Bang, drops to a negligible value that cannot be detected by astronomical observations.

The question naturally arises about the possible anisotropy of the Universe. But how exactly can it manifest itself in astronomical observations? Without an answer to this question, we do not have the opportunity to choose between anisotropic and isotropic cosmological models, while remaining within the framework of scientific theories \citep{p7}. 

It is well known that at present the Universe is inhomogeneous on small scales because of the growth of fluctuations and the creation of structures is an integral part of a real cosmological evolution. A homogeneous universe filled with matter is unstable. This was already shown by Lifshitz for the FLRW model in \citep{p4}. However, the gravitational Jeans instability \citep{p5} was known before this model, and even before general relativity itself. The initial small perturbations grew, forming an observable large-scale structure (LSS), including clusters, voids, and attractors. This structure causes collective large-scale motion of galaxies and other space objects. Simplistically, they can be considered as falling on overdensities, occurring against the Hubble expansion background. 

Cosmic flows and other manifestations of the LSS are actually observed. Unfortunately, sometimes these effects are mistaken for signature of the anisotropy of the Universe. One can find them in the list of alleged deviations from the FLRW model presented in \citep{p6}. We can add some more effects, which could erroneously be attributed to manifestations of anisotropy. For example, the radial motion of galaxies with a distance of up to 100 Mpc is characterized by the anisotropic Hubble parameter \citep{PP}. A statistically significant anisotropy is also observed in the orientations of galaxies \citep{PKK}. However, the first effect is associated with the influence of the Great Attractor, the Perseus-Pisces supercluster, the Shapley concentration and other overdensities. The second one is associated with the existence of the Local Cluster. So, these phenomena are caused by LSS.

The search for any deviations from the standard cosmological model, which is understood as FLRW $\Lambda$CDM model in which inhomogeneities do not arise and do not grow, contributes to the mixing impacts of anisotropy and inhomogeneity. Therefore, it is necessary to clearly separate the influence of these two different phenomena. Note that the requirement for homogeneity, i.e. the similarity of different points in space is often called the Copernican principle, the principle of mediocrity, or the cosmological principle \citep{p99}. Considering it to be certainly true, one can erroneously attribute any deviations from the standard model to the action of anisotropy. Deviations from uniformity are well known to astronomers. Baryon matter forms stars, galaxies and their clusters. Dark matter is concentrated in the halos of galaxies, which often contain supermassive black holes. On a large scale, we observe Laniakea and other LSS details \citep{DC}.

How could pure anisotropy manifest itself, if space-time is free from the influence of effects caused by inhomogeneity? An anisotropic homogeneous $\Lambda$CDM model is of little interest for studying the manifestations of anisotropy since they do not appear after the inflationary epoch. We need a completely different solution, which can be taken as a standard of anisotropic space-time. It must preserve the degree of anisotropy (a quantitative value expressing it was introduced in \citep{p3}). Therefore, in this standard model, the universe should contain neither matter nor dark energy, so that isotropization does not occur. So we must choose some vacuum model without a cosmological constant. To simplify the study, we can consider a homogeneous cosmological model, preferably a simple one. As a result, the Kasner cosmological model \citep{p8} is the best choice for this reference model. The properties of this solution are described in Section \ref{s3}.

The Kasner metric stands out not only for its simplicity. Many exact solutions of the Einstein equations are simple generalizations of this metric near their singularities \citep{p1}. The generalized Kasner solution near cosmological singularity was studied in \citep{p10}. In more complex cases, for example, for homogeneous spaces of Bianchi types IX and VIII, the oscillatory solution near the singularity consists of an infinite number of epochs, during which the space-time is well described by the generalized Kasner solution \citep{p11}. The presence of matter or the cosmological constant changes the form of the solution, but not its Kasner asymptotics (see for example \citep{p9}).

All of the above draws special attention to the properties of the space-time described by the Kasner metric. How well do we understand the meaning of the Kasner's metric? Here is what is written about this in \citep{p9}: ``the metric corresponds to a homogeneous but anisotropic space whose total volume increases (with increasing $t$) proportionally to $t$; the linear distances along two of the axes ($y$ and $z$) increase, while they decrease along the third axis ($x$).'' Let us show that this statement, although true, does not give a complete idea of the properties of space-time with the Kasner metric. We study them in this article.

Imagine an astronomer in a world with the Kasner metric. What they would see in a telescope? Would objects be shifted relative to their true position in the sky? Would they move? What about the redshift of their spectra? What manifestations of the observed space-time anisotropy could astronomer discover in the course of their observations? The answer is not as obvious as it may seem at first sight. 

One detail is clear initially. An astronomer can only observe objects that are closer than the cosmological horizon. This follows from the fact that both the age of the universe at the moment of observation and the speed of light are finite. The shape of the horizon is discussed below, and its size increases with time. At the very horizon, an astronomer could observe radiation emitted at the time of the anisotropic Big Bang. More distant objects are not visible until they get inside the expanding cosmological horizon.

Let's make one clarification to be puristic. Astronomers do not see the Big Bang because of the real Universe was filled with opaque plasma before the era of recombination. They see the so-called last scattering surface instead of it. It delineates the edge of the observable Universe. But matter is absent in our speculative task and the universe is transparent at all stages of its evolution after the annihilation of electron-positron pairs. We are not interested in discussing the issue of the appearance of such pairs at the earliest stages of the existence of a universe with the Kasner metric. So we simply assume that it has been transparent since the Big Bang.

The Kasner metric poorly describes the real Universe. However, we study it to determine observational manifestations of anisotropy. To do this, we study the behavior of light geodesics in the Kasner space in Section \ref{s3}. At the same time, we can better understand the properties of this solution. After searching for observed manifestations of anisotropy, we consider in Section \ref{s4} whether they are observed by astronomers. General conclusions are described in Section \ref{s5}.

\section{Geodesic structure and properties of the Kasner's space-time}\label{s3}
\subsection{The Kasner metric}\label{s2}
This vacuum solution was obtained in 1921 \citep{p8} and can be considered as a reference anisotropic cosmological model. The space in this case is homogeneous and belongs to the Bianchi type I. The metric has the form
\begin{equation}
\label{eq1}
\rmd s^{2} = \rmd t^{2} - t^{2p_{1}} \rmd x^{2} - t^{2p_{2}} \rmd y^{2} - t^{2p_{3}}\rmd z^{2}.
\end{equation}
We use a system of units in which the speed of light is equal to one ($c=1$). The coordinate $x^0=t$ is the cosmological time. We denoted spatial coordinates by letters $x^i=(x,y,z)$. The Latin indices run from 1 to 3 (e.g. $i=1,2,3$), while the Greek ones change from 0 to 3. All other designations and signs coincide with those used in the book \citep{p9}. 
The set of constants $p_i$ are called Kasner indices. They satisfy two conditions, namely
\begin{equation}
\label{eq2}
p_{1} + p_{2} + p_{3} = 1,\quad {p_{1}}^{2}
+{p_{2}}^{2} +{p_{3}}^{2} = 1.
\end{equation}
They can be expressed in terms of a single parameter $w$
\begin{equation}
\label{eq3}
p_{1} =-\frac{w}{1+w+w^2},\quad p_{2} =\frac{1+w}{1+w+w^2},\quad p_{3} =\frac{w+w^2}{1+w+w^2}.
\end{equation}
One of the indices must lie in the range from -1/3 to 0, the second one from 0 to 2/3, the third one from 2/3 to 1.  

Space-time (\ref{eq1}) has a cosmological space-like singularity at $t=0$. The invariants of the curvature tensor diverge in it, unless one of the Kasner indices is equal to 1, and the other two are equal to 0, then this is part of the Minkowski space-time.

The metric (\ref{eq1}) is simple and depends on only one variable. The coordinate system is synchronous. Three Killing vectors are directed along the $x^i$ coordinate axes. The space is also invariant under mirror reflections with respect to $x^i=0$ planes.

Although the metric (\ref{eq1}) describes a vacuum solution, we can consider a motion of test particles, which create a negligibly weak gravitational field. Very light motionless particles can be used as such indicators of motion. They remain motionless in the $x^i$ coordinate system due to the homogeneity of space. The components of their 4-velocity $u^\alpha$ are (1,0,0,0). We consider one of these test particles as the location of the observer, and the other as the location of the celestial object, the light from which reaches the observer.
\subsection{Null geodesics}
Let's use the advantages of the coordinate system $x^\alpha$ in order to find null geodesics in space-time (\ref{eq1}). We use the eikonal equation
\begin{equation}
\label{eq4}
g^{\alpha\beta}\frac{\partial\Psi}{\partial x^{\alpha}}\frac{\partial\Psi}{\partial x^{\beta}}=\left( \frac{\partial\Psi}{\partial t}\right)^2 - t^{-2p_1}\left( \frac{\partial\Psi}{\partial x}\right)^2 - t^{-2p_2}\left( \frac{\partial\Psi}{\partial y}\right)^2 - t^{-2p_3}\left( \frac{\partial\Psi}{\partial z}\right)^2=0.
\end{equation}
The variables are easily separated 
\begin{equation}
\label{eq5}
\Psi=k_xx+k_yy+k_zz+\Psi_t(t),\quad k_x,k_y,k_z=\rm{const}
\end{equation}
and we get
\begin{equation}
\label{eq6}
\Psi_t(t)=\pm\int \left(t^{-2p_1}k_x^2+t^{-2p_2}k_y^2+t^{-2p_3}k_z^2\right)^{1/2} d\rm{t}.
\end{equation}
The trajectories of light rays are obtained from the conditions
\begin{equation}
\label{eq7}
\frac{\partial\Psi}{\partial k_x}=\rm{const},\,\frac{\partial\Psi}{\partial k_y}=\rm{const},\,\frac{\partial\Psi}{\partial k_z}=\rm{const}.
\end{equation}
Coordinates are expressed as functions of time
\begin{eqnarray}
\label{eq8}
x=\pm k_x \int \frac{t^{-2p_1}}{\sqrt{k_x^2 t^{-2p_1}+{k_y^2 t^{-2p_2}+{k_z^2 t^{-2p_3}}}}} \rmd t , \nonumber \\ y=\pm k_y \int \frac{t^{-2p_2}}{\sqrt{k_x^2 t^{-2p_1}+{k_y^2 t^{-2p_2}+{k_z^2 t^{-2p_3}}}}} \rmd t,\nonumber \\
z=\pm k_x \int \frac{t^{-2p_3}}{\sqrt{k_x^2 t^{-2p_1}+{k_y^2 t^{-2p_2}+{k_z^2 t^{-2p_3}}}}} \rmd t.
\end{eqnarray}
This gives the equation of null geodesics in the parametric form. 
\subsection{Features of the $x,y,z$ coordinate system}
Before analyzing these geodesics, we pay attention to some features of the spatial coordinate system $x^i$. Let's start with the dimension. A system of units is used in which time is measured in seconds and length in (light) seconds. The dimension of coordinates $x^i$ includes a very unusual combination. This is the time to the power of $p_i$. The reason for this becomes clear if we understand that the spatial part of the metric (\ref{eq1}) should include not $t^{p_i}$, but the dimensionless ratio of $t$ to $t_0$ to the power of $p_i$. Here $t_0$ is some constant with the dimension of time. However, scaling of the coordinates $x^i$ was used when deriving the Kasner metric in the paper \citep{p8}. As a result, the parameter $t_0$ disappeared from the metric, but the coordinates $x,y,z$ received non-trivial dimensions.

The notation of these coordinates is similar to Cartesian and they are mutually orthogonal. However, this system should not be compared with the Cartesian one. This is a kind of generalization of the concept of curvilinear orthogonal coordinates for the case of curved space. The analogues of Lam\'e coefficients for them depend on time and are equal to $t^{p_i}$.

Accordingly, the meaning of the constants, which we have designated as $k_x,k_y,k_z$, is not so obvious. There may be a false impression that these are the lengths of  projections of a wave vector onto three mutually orthogonal axes. We can introduce the concept of the vector $\vec k$. Its projections on the $x^i$ axis $k_x,k_y,k_z$ are constant. But this is not a wave vector of the electromagnetic wave from a source to an observer. The vector $\vec k$ is constant, while the wave propagates along a curved trajectory.

Let's demonstrate it. Consider the case $k_z=0$. Such null geodesic passes in the $xy$ ``plane''. We denote $\beta \equiv k_y/k_x=\rm{const}$ and the angle between the direction of the geodesic and the $x$-axis as $\phi$. In the time interval $dt$ the coordinate $x$ changes by the value $dx=k_x t^{-2p_1}(k_x^2 t^{-2p_1}+k_y^2 t^{-2p_2})^{-1/2} dt$ and the coordinate $y$ by $dy=k_y t^{-2p_2}(k_x^2 t^{-2p_1}+k_y^2 t^{-2p_2})^{-1/2} dt$. The shifts of the geodesic along the axes $x$ and $y$ are $t^{p_1}dx$ and $t^{p_2}dy$, respectively. Its slope $\tan \phi$ in the $xy$ ``plane'' is calculated as rise over run. Therefore 
\begin{equation}
\label{eq9}
\tan \phi=\beta t^{p_1-p_2}.
\end{equation}
We see that the angle $\phi$ changes with time. This means the deviation of the wave vector from its original direction. Null geodesics are not straight lines, but curves. It is just as easy to find time-like geodesics by separating the variables in the Hamilton-Jacobi equation. They are also curved.
\subsection{Chosen ``astronomical'' coordinate system}
But hypothetical astronomer would not use the $(x,y,z)$ spatial coordinate system. The natural choice of the coordinate system is different one. The curvature invariants of the Kasner metric drop to zero at $t\to\infty$ and the space-time (\ref{eq1}) is asymptotically flat. However, the metric (\ref{eq1}) does not tend to the Minkowski one at $t\to\infty$. It happens in the coordinate system in which the cosmological time $t$ is supplemented with a set of three spatial coordinates which have the dimension of length
\begin{equation}
\label{eq10}
\xi^i= t^{p_i} x^i.
\end{equation}
Here and below there is no summation over repeated indices. The metric (\ref{eq1}) takes the form
\begin{equation}
\label{eq11}
\rmd s^{2} = \rmd t^{2} - \sum_{i=1}^3 \left(\rmd \xi^i-p_i\xi^i t^{-1}\rmd t\right) ^2.
\end{equation}
It is a natural system for an observer studying asymptotically flat space-time at $t\to\infty$. Let's call it conditionally ``astronomical''.

The distances in this coordinate system are described by the spatial metric
\begin{equation}
\label{eq12}
\gamma_{ik}=-g_{ik}+\frac{g_{0i}g_{0k}}{g_{00}}=\delta_{ik}+\frac{p_ip_k\xi^i\xi^k}{t^2-\sum_{j=1}^3(p_j\xi^j)^2},
\end{equation}
where $\delta_{ik}$ is the Kronecker symbol. These distances are determined by location, i.e. by the time of propagation of light to the object and its return to the starting point. This is one of the most accurate methods used in real measurements of distances to celestial bodies. It is clear that it is used to determine the distance to nearby objects, for example, in laser ranging of the Moon and planets, where the influence of the expansion of the Universe can be neglected. But it is possible to determine the distance to faraway objects using a location method, at least in a gedanken (thought) experiment.

In real astronomical observations, the distances to the nearest stars were determined from their parallaxes, i.e. from their displacement on the celestial sphere when comparing observations with an interval of six months. In this case, the astronomer is in opposite parts of the Earth's orbit. The parallax of an object is equal to the angle under which this orbit is visible from the location of the object. In general relativity, this method determines the so-called distance from angular size. Further in the article, we show that in the astronomical coordinate system for nearby celestial objects, such distances approximately coincide with the radar distances obtained by location i.e. from (\ref{eq11}). 

Let us clarify the term ``nearby'' here. The metric (\ref{eq1}) by itself does not have a characteristic temporal or spatial scale. However, it appears when the problem is correctly formulated. The time scale is determined by the age of the universe, which we denote by the letter $T$. It determines the characteristic spatial scale, namely the size of the cosmological horizon. We consider an object being close if the distance to it $R$ is significantly less than this size of this horizon, $R\ll T$.

Having introduced all these concepts and definitions, we can begin to study the trajectories of light rays in the Kasner space-time (1) both in the coordinates $x^i$ and in the astronomical coordinate system $\xi^i$. The orientation of the axes is fixed and given by the properties of space-time (\ref{eq1}). We can consider rays emitted by a point source and propagating along the light cone of the future, or rays of light that have come along the light cone of the past to the observer's location. These cases differ only in the direction of time $t$. In both cases we can take advantage of the homogeneity of space and place the origin of coordinates $x^i$ and $\xi^i$ at the point where the source or observer is located. We assume that this point is fixed in the comoving coordinate system $x^i$ like any other test body. We use the equations of motion (\ref{eq8}) and the connection of two coordinate systems (\ref{eq10}). 
\subsection{Geodesics in the direction of the axes}
Let's consider some particular cases. If two of the three constants $p_x,p_y,p_z$ are equal to zero, then the light propagates along one of the axes with the Kasner index $p_i$. Coordinates $x^i$ and $\xi^i$ have changed its value during the time $T$ elapsed since the Big Bang by 
\begin{equation}
\label{eq13}
\Delta x^i=\int_0^T t^{-p_i} \rmd t=\frac{T^{1-p_i}}{1-p_i},\quad\Delta \xi^i=\Delta x^iT^{p_i}=\frac{T}{1-p_i}.
\end{equation}
These quantities are the coordinates of the intersection points of the axes with the cosmological horizon. We see that this horizon is not a sphere. Its shape resembles a triaxial ellipsoid with semiaxes $T(1-p_1)^{-1}, T(1-p_2)^{-1},T(1-p_3)^{-1}$ in $\xi^i$ coordinates. The distance to the horizon is less than $T$ along the axis with a negative Kasner index. It exceeds $3T/4$. The horizon coordinate is greater than $T$ along the other two axes. It lies in the interval from $T$ to $3T$ for an axis with an middle index. It is greater than $3T$ and can become arbitrarily large as $p_i \to 1$ for the axis with the maximum index. However, at $p_i =1$ the singularity in the metric (\ref{eq1}) is fictitious and it describes a part of the Minkowski space-time. Therefore, the distance to the cosmological horizon cannot become infinitely large.

Let us consider light emitted by a nearby object with coordinate $\xi^i=l\ll T$ and propagating along the axis with index $p_i$ to the origin of coordinates. The distance to this object can be determined from (\ref{eq12}). It is approximately equal to
\begin{equation}
\label{eq14}
L\approx \int_0^l \sqrt{1+p_i^{\phantom{i}2}(\xi^i)^2T^{-2}}\,\rmd \xi^i\approx l+\frac{p_i^{\phantom{i}2}}{6T^2}l^3\xrightarrow[T\to\infty]{}l.
\end{equation}
The object moves along the axis. This can be seen both from the change in $L$ with time and from the fact that a test body immobile in the $x^{\alpha}$ coordinate system after the transition to the system $(t,\xi^i)$ has a four-dimensional velocity
\begin{equation}
\label{eq15}
u^{\prime\,\alpha}=\left(1,\frac{\partial \xi^i}{\partial t}\right)=\left(1,\frac{p_i\xi^i}{T}\right).
\end{equation}
The  recession velocity is approximately equal to $V\approx p_iT^{-1}l$. So, we can introduce something like an anisotropic Hubble parameter for close objects with $l\ll T$. It is approximately equal to $p_i/T$ along the $\xi^i$-th axis. It is positive along two axes and negative along the third and decreases with time. 

However, the issue of redshift requires special consideration. The observer can measure the shift of the spectra of nearby test bodies. Is it caused by their radial motion? Let's formulate the problem. The observer sees the light from the source at the moment when the age of the universe is $T$ with a delay $\tau(T)$ after its emission. Light propagates along the $x^i$-axis. The source and the observer are fixed in the coordinate system $x^i$, in which the distance between them is equal to $\Delta x^i$. From equation (\ref{eq8}) we get the condition
\begin{equation}
\label{eq16}
\Delta x^i=\int_{T-\tau}^T t^{-p_i} \rmd t=\frac{T^{1-p_i}-(T-\tau)^{1-p_i}}{1-p_i}=\rm{const}.
\end{equation}
We get $\tau\propto T^{p_i}$ at $\tau\ll T$. It is easy to understand that the frequency of the observed light $\nu$ is related to the frequency of the emitted light $\nu_0$ by the relation
\begin{equation}
\label{eq17}
\nu=\nu_0\left(1+\frac{\partial \tau}{\partial T}\right)^{-1},\quad \frac{\nu_0-\nu}{\nu_0}\approx \frac{p_i}{T}\tau.
\end{equation}
This corresponds to a Doppler shift when moving at a speed $V$.
\subsection{Geodesics on surfaces passing through two axes}
Consider geodesics for which one of the quantities $k_x,k_y,k_z$ is equal to zero. For definiteness, we set $k_z=0$. The geodesics lie on the surface passing through the $x$ and $y$ axes. If space were flat, we would call it the $xy$ plane. If light is emitted by a source located at a point with coordinates $x=x_1,y=y_1,z=0$ and at time $t=T$ is observed at a point with coordinates $x=x_2,y=y_2,z=0$, then from equations \eqref{eq8} we obtain two conditions for this case
\begin{eqnarray}
\label{eq18}
\Delta x\equiv x_2-x_1=\int_{T-\tau}^T \frac{t^{-p_1}}{\sqrt{1+{\beta^2 t^{2(p_1-p_2)}}}} \rmd t
\nonumber \\=\frac{t^{1-p_1}}{1-p_1}{}_{2}F_1\left(\frac{1}{2},\frac{1-p_1}{2(p_1-p_2)};\frac{1-p_1}{2(p_1-p_2)}+1;-\beta^2t^{2(p_1-p_2)}\right)\bigg|^T_{T-\tau},\nonumber \\ \Delta y\equiv y_2-y_1=\beta\int_{T-\tau}^T \frac{t^{p_1-2p_2}}{\sqrt{1+{\beta^2 t^{2(p_1-p_2)}}}} \rmd t\nonumber \\=\frac{\beta t^{1+p_1-2p_2}}{1+p_1-2p_2}{}_{2}F_1\left(\frac{1}{2},\frac{1+p_1-2p_2}{2(p_1-p_2)};\frac{1+3p_1-4p_2}{2(p_1-p_2)};-\beta^2t^{2(p_1-p_2)}\right)\bigg|^T_{T-\tau}.
\end{eqnarray}
Here $\Delta x,\Delta y$ are constants, ${}_2F_1$ is the hypergeometric function and $\beta=p_y/p_x=\rm{const}$. The latter, like the delay time of the signal $\tau$, depends on $T$. These two unknown functions could be found from two conditions \eqref{eq18}. 

What effects are hidden behind the dependencies of $\beta$ and $\tau$ on $T$? The first shows that null geodesics do not simply bend, which causes the apparent position of objects on the sky to deviate from their true position. The deflection angle changes with time, i.e. objects move across the sky, showing non-existent proper motion. Its speed is determined both by the position of the object in the sky and the distance to it. This  proper motion occurs in opposite directions for objects that are symmetrical about an axis. Therefore, it is difficult to take it for a general rotation.

The dependence of $\tau$ on $T$ means not only the change in the delay time, but also the redshift or blueshift associated with it in accordance with the formula (\ref{eq17}). As shown above, there is a red shift near the axes with positive Kasner indices, and blue shift near the axis with a negative index. Over time, space-time (\ref{eq11}) tends to the Minkowski metric. Therefore, deflection angles, proper motion velocities and redshift decrease and tend to zero.

Note that these conclusions are drawn from general considerations. A specific type of conditions (\ref{eq18}) may be needed only in order to exclude the option in which either $\beta$ or $\tau$ do not depend on $T$. Checking this possibilities is reduced to long and not very interesting transformations of formulas. We omit an analysis of these impossible cases.
\subsection{Distance from angular size for nearby sources}
We need the equations (\ref{eq18}) to check that distances from angular size are close to radar distances for nearby objects in the Kasner space-time (\ref{eq1}) when using an astronomical coordinate system. From (\ref{eq14}) we know that the latter tends to the $x$-coordinate of the beam directed along the $x$-axis at $T\to\infty$. Let the light source be deflected from this axis and have coordinates $(\xi^1,\xi^2,0)$ in the astronomical coordinate system. In flat space, a ray propagates along a straight line inclined at an angle $\gamma=\tan^{-1}(\xi^2/\xi^1)$ and reach at the time $t=T$ the origin of $\xi^i$ coordinates where the observer is located. At what angle to the axis does it come to an observer in the Kasner's space-time?

The relationship between the coordinates of a point on a beam of light in $x^i$ coordinate systems and astronomical one near the origin of coordinates takes the form
\begin{equation}
\label{eq19}
d(\xi^1)=d(xt^{p_1})=t^{p_1}dx+p_1xt^{p_1-1}dt\to T^{p_1}dx
\end{equation}
due to $t\to T$ and $x\to 0$. Similarly $d(\xi^2) \to T^{p_2}dy$. We get
\begin{eqnarray}
\label{eq20}
\tan \gamma =\frac{d(\xi^2)}{d(\xi^1)}\to T^{p_2-p_1}\frac{dy}{dx}=T^{p_2-p_1}\frac{dy/d\tau}{dx/d\tau}\nonumber \\
=T^{p_2-p_1}\frac{\beta T^{p_1-2p_2}\left(1+\beta^2T^{2(p_1-p_2)}\right)^{-1/2}}{T^{-p_1}\left(1+\beta^2T^{2(p_1-p_2)}\right)^{-1/2}}=\beta T^{p_1-p_2} \to \tan \phi.
\end{eqnarray}
We used the equation (\ref{eq9}). So, for nearby objects, both the radar and angular distances tend to the distance measured in Minkowski space, to which solution (\ref{eq1}) tends in the limit at $t\to \infty$. This confirms that the astronomical coordinate system is distinguished. Note that error in the radar distance measurement compared to flat space-time falls as $T^{-2}$ according to (\ref{eq14}) and the error in the distance from angular size is proportional to $T^{-1}$.

The anisotropy manifests itself in the shift of the emission spectrum, which can correspond to both the redshift and the blueshift, and in the apparent proper motion. Both of these effects decrease with time and increase with increasing distance from the object. They cannot be overlooked for distant objects.
\subsection{Light rays from distant sources}
Let us consider null geodesics propagating along the surfaces $x=0$, $y=0$, and $z=0$, which move away far enough from the source. In other words, we study the light emitted by objects located not very far from the cosmological horizon. Naturally, we use equations (\ref{eq18}) for rays with $z=0$. We choose the origin of coordinates at the observer's location ($x_2=y_2=0$) and reduce the number of parameters in the integrals as much as possible. To do this, we choose the age of the universe $T$ as the natural time scale and express the quantities $t$ and $\tau$ in its fractions. We introduce two dimensionless parameters $\eta=t/T$ and $\varepsilon=\tau /T$ and denote $A=\beta T^{p_1-p_2}$. We can rewrite (\ref{eq18}) in the form
\begin{equation}
\label{eq21}
x(\varepsilon)=T^{1-p_1}\int_{1-\varepsilon}^1 \frac{\eta^{-p_1}\rmd \eta}{\sqrt{1+{A^2 \eta^{2(p_1-p_2)}}}} ,\quad y(\varepsilon)=AT^{1-p_2}\int_{1-\varepsilon}^1 \frac{\eta^{p_1-2p_2}\rmd \eta}{\sqrt{1+{A^2 \eta^{2(p_1-p_2)}}}} .
\end{equation}
We can go over to dimensionless quantities $X=xT^{p_1-1}$ and $Y=yT^{p_2-1}$. They are expressed in terms of integrals whose integrand depends on $A$, and the lower limit is expressed in terms of the parameter $\varepsilon$. The value of $A$ is directly related to the angle between the geodesic and the $x$ axis at the point of observation. If the values $X$ and $Y$ are plotted along the axes on the same scale, then the tangent of this angle is equal to $A$. In the figures below we show the trajectories of five rays for which this angle is equal to $15^{\circ}$, $30^{\circ}$, $45^{\circ}$, $60^{\circ}$, and $75^{ \circ}$, respectively. The dark circles mark the points at which the parameter $\varepsilon$ takes the values 0.1, 0.2, 0.3, 0.4, 0.5, 0.6, 0.7, 0.8 and 0.9. All geodesics reach the origin at $\varepsilon=0$. The edges of the trajectories farthest from the observer correspond to $\varepsilon=0.999$, which is close to the cosmological horizon at $\varepsilon=1$. 

The figures show some of the trajectories located in the first quadrant of coordinates on the surface $x=0$, $y=0$, or $z=0$. The complete picture is obtained after adding a mirror reflection of the trajectories about the axes and ones that are symmetrical about the origin of coordinates.

The metric (\ref{eq1}) remains the same when the order of coordinates $x^i$ is changed. For definiteness, we choose such an order in which $-1/3<p_1<0<p_2<2/3<p_3<1$. It corresponds to the case $w>1$ in the equation (\ref{eq3}). It is in this order that we  choose the coordinates $x^i$ in the figures with which we illustrate the behavior of the light rays. They are calculated for a set of Kasner indices (-2/7, 3/7, 6/7) corresponding to $w=2$. The behavior of the trajectories does not change qualitatively for any set of indices.

\begin{figure}\begin{center}
\includegraphics[width=12cm]{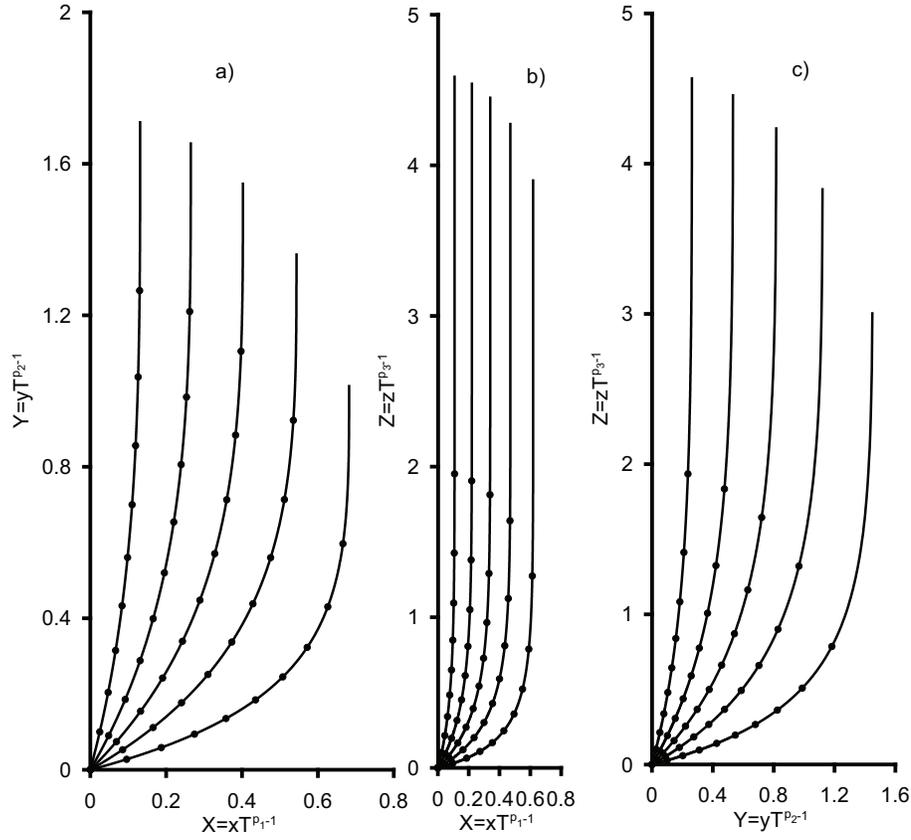}
\end{center}
\caption{\label{f1}Null geodesics on surfaces passing through two axes of the coordinate system. Surfaces $z=0$, $y=0$ and $x=0$ correspond to panels a), b) and c). The quantities $X$, $Y$, and $Z$ are plotted on the axes. The dark circles correspond to the propagation time of light, which is a multiple of 1/10 of the age of the Universe. Far ends of the lines correspond to $\varepsilon=0.999$ near the cosmological horizon.}
\end{figure}
\begin{figure}\begin{center}
\includegraphics[width=12cm]{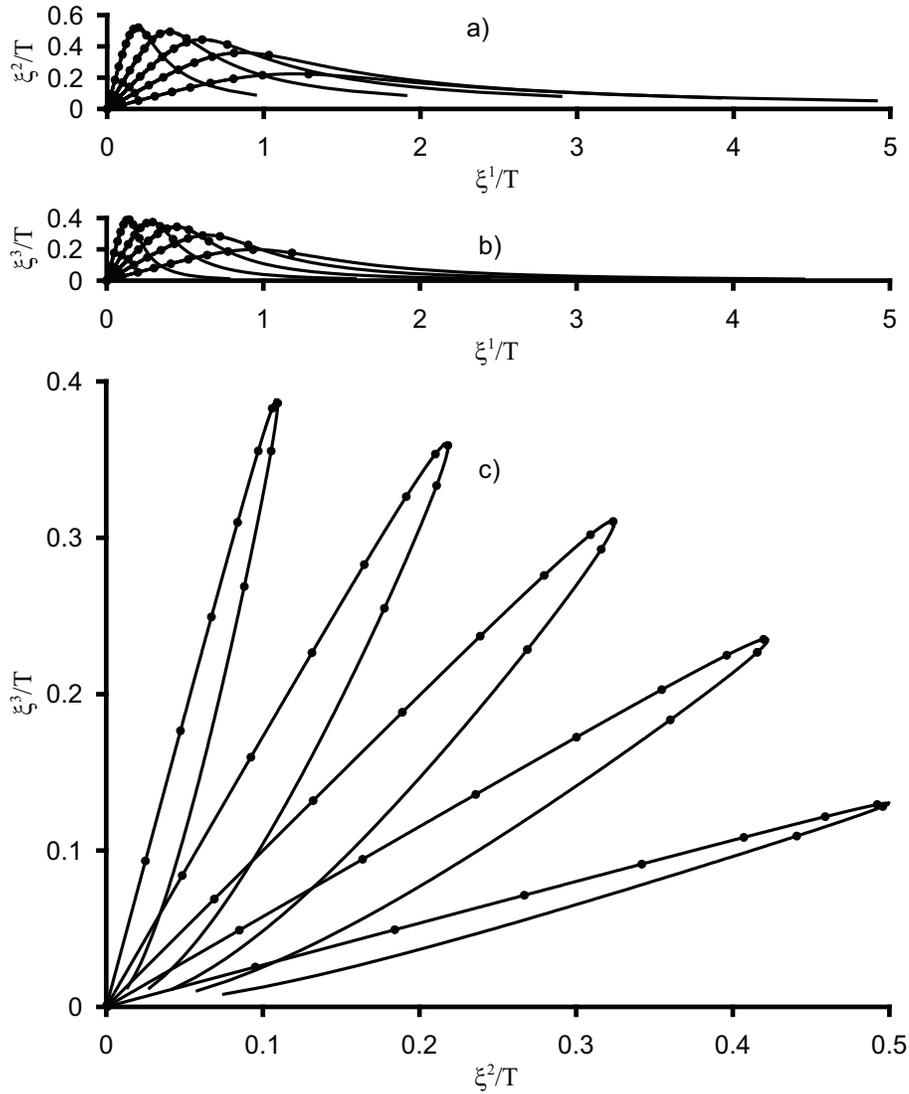}
\end{center}
\caption{\label{f2}The same as in Figure \ref{f1}, but in ``astronomical'' coordinates $\xi^i$.}
\end{figure}

Figure \ref{f1} shows geodesics on the surface $z=0$ in the coordinates $X.Y$ (panel a), on the surface $y=0$ in the coordinates $X,Z=zT^{p_3-1}$ (panel b), and on the surface $x=0$ in the coordinates $Y,Z$ (panel c). The same geodesics are shown in $\xi^i$ coordinates in the Figure \ref{f2}.

What conclusions can be drawn from these figures? First of all, it is a confirmation that the system of astronomical coordinates $\xi^i$ is very convenient for describing nearby objects. In our example, we can see that we can consider objects with $\varepsilon<0.6$ as nearby ones. Geodesics in this case are close to straight lines, which make it possible to determine distances from parallaxes. Light rays in $X,Y,Z$ coordinates are more curved, but the angle between the true and apparent position in the sky is not very large. The difference between these two coordinate systems is mainly due to the fact that in the comoving coordinate system $x^i$ the light sources are stationary, while in the astronomical system they have a decreasing radial velocity component. The observed red and blue shifts in one case can be described as a Doppler shifts, in the second as a consequence of the change in the time of light passage from the source to the observer in accordance with formula (\ref{eq17}).

But none of the coordinate systems gives a good description of the observed sky pattern for large distances. Theories and methods that work satisfactorily for nearby objects become inadequate for distant ones. The use of astronomical coordinates $\xi^i$ becomes impossible for describing distant objects. As $t$ increases, the two coordinates $\xi^2$ and $\xi^3$ initially increase, and then (when propagating not along the $x$ axis) reach a maximum and decrease. 

The geodesics in Figure \ref{f2} intersect. It may seem that the light emitted by the source can come to the observer in two different ways, and we are dealing with a special case of gravitational lensing. But this is not so, since distant sources change their position in astronomical coordinates quite quickly, and two geodetic ones in Fig. \ref{f2} pass the intersection point at significantly different times. The geodesics in the comoving coordinate system do not intersect in Figure \ref{f1}. This confirms the absence of gravitational lensing.
\subsection{Visible manifestations of space anisotropy for distant objects}
We can calculate the shape of a ray emitted in an arbitrary direction with the help of formulas (\ref{eq8}). What are the effects of going beyond surfaces that pass through two axes? Obviously, the displacement of objects in the sky and their apparent proper motion  changes both angles that describe positions on the celestial sphere, such as right ascension and declination.

There is another group of effects caused by the curvature of the geodesics in Figure \ref{f1}. For example, if light sources are evenly distributed over the celestial sphere, then the observed position of nearby sources is fairly isotropic. However, the distant ones are clearly concentrated in the direction of the axis with a negative Kasner index and are rarefied near the axis with the maximum one. 

The curvature of light rays in space-time (\ref{eq1}) leads to two more effects, which are observed more clearly for distant objects. The first is the distortion of images by shear deformation. It is not local, as in gravitational lensing, but global, covering the entire celestial sphere.

The second is related to the distortion of photometric distances to radiation sources. We cannot determine the distance to far objects using location or their parallax. But we can do this by the magnitude of the electromagnetic radiation flux observed on Earth. It falls in inverse proportion to the distance to the source, in accordance with the law discovered by Kepler in the early 17th century. This makes it possible to obtain the so-called photometric distance to the source. It underlies many methods for estimating the distance from Cepheids to the light curves of Ia supernova outbursts. In real astronomy, in addition to the three listed methods for determining the distance, an estimate is also used by the redshift, but in our gedanken experiment in space with metric (\ref{eq1}) the latter does not work.

Naturally, Kepler's formula does not work well with a significant extinction of the medium that fills the intergalactic or interstellar space and with a significant influence of the space-time curvature. It is with the latter that we are dealing in the case under study. If the radiation in a direction near the $x$ axis with a negative Kasner index lies in a small solid angle $d\Omega$, its cross-sectional area at a distance $R$ increases faster than $R^2d\Omega$. As a result, the  illuminance or radiation flux falls faster and the calculated photometric distance is greater than the real one. On the other hand, in the absence of extinction, the law of conservation of energy ensures the existence of directions along which the radiation flux falls weaker than according to the Kepler law, and the estimate of the photometric distance becomes less than the real one. These include, in particular, the direction of the $z$-axis with the maximum Kasner index.

In our case, when sources uniformly distributed in space, the observer discovers a direction-dependent distribution of their photometric distances. Note that the anisotropy axes in studying the distributions of photometric distances and the surface density of objects inside spherical layers must be the same. Indeed, the overestimation or underestimation of photometric distances is a consequence of the variation in the density of the apparent distribution of sources over the sky. Therefore, these two manifestations of anisotropy are closely related.

More precisely, the distribution of radiation fluxes from objects is determined by the distribution of distances to them and the effect under discussion. As noted above, the cosmological horizon is closest to the astronomer in the direction of the $x$-axis with a negative Kasner index. The distance to it $T/(1-p_1)$ lies in the range from $0.75T$ to $T$. The same distance along the $z$-axis with the maximum index exceeds $3T$. These distances differ by more than three times, and the corresponding fluxes from identical sources near the cosmological horizon differ by more than nine times. When estimating, we do not take into account the spectrum of the source and the effect of red or blue shift on the flux. This difference is partly compensated by the above mentioned difference in the density of the light rays due to their deflection.

The simplest manifestations of the anisotropy of space-time can be the anisotropy of the distribution of the density of objects over the celestial sphere and their luminosity, taken for very distant sources, the distance to which lies within a given interval. These indicators should be anticorrelated.
\subsection{How the degree of anisotropy is maintained}
Note that the pattern of geodesics in Figures \ref{f1} and \ref{f2} does not change with the age of the universe $T$, except for changing the scale of the axes. Therefore, it is qualitatively independent of time. This statement needs to be clarified. An astronomer observes radiation from specific sources. For any of them, as the age of the universe $T$ increases, change both the signal propagation time $\tau$ and the value of the derivative $\partial\tau /\partial T$, which determines the shift of the spectrum according to (\ref{eq17}). Consequently, the redshift of each of the objects changes with time, asymptotically tending to zero.

Let us consider a specific source at some point in time $t=T_1$, determine its apparent position in the sky and its redshift. Then, at any subsequent time $t=T_2>T_1$, we consider a beam of light coming from this direction and mark on it the point at which the parameter $\varepsilon$ has the same value as for the first object at the time $t=T_1$ and place the second, conditional source of radiation in it. For these two objects, the initial one at $t=T_1$ and the second one at $t=T_2$, all the observed parameters coincide: position in the sky, light propagation time, and redshift. If we consider all the observed objects, then at time $t=T_2$ there will be more of them than at $t=T_1$ due to the fact that when the cosmological horizon expands, new objects will fall inside it. But their distribution over the apparent position in the sky, over the parameter $\varepsilon$, over the redshift, and over any combination of these parameters will not change.

We noted that the distributions of the density of observed objects and their photometric distances over the sky can serve as indicators of anisotropy. We can consider relative values by dividing these distributions by the total number of observed objects. These normalized distributions are constant in time. The emission from each source becomes less and less anisotropic over time, but new sources emerging from beyond the cosmological horizon keep these normalized distributions constant. This statement assumes the constancy of the volume density of the number of radiation sources, i.e. homogeneity of their distribution. This is a fairly natural assumption for a homogeneous cosmological model. But real galaxies or quasars do not exist forever, starting from the moment of the Big Bang. It takes time for their formation and the appearance of radiation sources, such as stars in galaxies. This may lead to a change in time of the discussed normalized distributions, but not to the disappearance of the observed manifestations of anisotropy.
\section{No observational manifestations of space-time anisotropy}\label{s4}
We found out what are the possible manifestations of the anisotropy of space on the example of solution (\ref{eq1}). They are especially pronounced when observing distant objects. But nothing like this is seen in real astronomical observations. The most distant of the observed radiation is the very isotropic CMB. It was emitted at a time when the Universe was only 379,000 years old. Dividing by its current age, we get $1-\varepsilon=2.7\cdot 10^{-5}$. No anisotropy was found in the distribution of the earliest galaxies with $z\gtrsim 10$. From this we can conclude that the Universe since the moment of recombination is very isotropic. Cosmic flows and other phenomena are manifestations of the LSS. The latter, in turn, is the result of a growth of fluctuations due to gravitational instability.

Note a feature of the Kasner indices associated with conditions (\ref{eq2}). They allow for the possibility that the two indices are close in magnitude. This is a set of exponents close to (-1/3, 2/3, 2/3) and (0,0,1), while the set (0,0,1) itself is not considered due to the fictitious Kasner singularity in this case. But all three indices cannot become equal or close in value. Therefore, observational manifestations of anisotropy in space-time (\ref{eq1}) cannot be weak.

What manifestations of anisotropy can be detected? This is primarily anisotropy in the density of object's distribution, especially the most distant ones. It is also possible to consider the anisotropy in the distribution of the apparent magnitude of objects, which is not associated with extinction in the Galaxy. It is associated with the previous one. There are no observable remoteness indicators other than redshift and apparent magnitude for the most distant objects. Therefore, it is more correct to speak about the deviation of the dependence between these two observed values for objects observed in different directions.

We can estimate the apparent proper motion of objects. The angle of deviation could change by a value of the order of ten degrees during the existence of the Universe. Proper motion for objects near the cosmological horizon is greater than for nearby ones. If, with increasing accuracy of astrometric observations, it turns out that very distant galaxies demonstrate proper motion with these parameters, then this would be evidence in favor of anisotropy. If this motion would be approximately symmetrical with respect to a certain plane, this would be a significant argument in favor of the existence of anisotropy.

In any case, manifestations of anisotropy should be sought primarily in distant galaxies and quasars, as well as in the CMB distribution over the sky. We can look for global anisotropy in the distribution of the CMB radiation flow coming from different directions. For distant galaxies and quasars, we can look for anisotropy in the distribution of luminosities of sources whose redshift lies in a certain interval and in the density of their distribution over the sky.
\section{Conclusions}\label{Conclusions}\label{s5}
We study the properties of the Kasner's space-time (\ref{eq1}) for a better understanding of its physical meaning. Particular attention is paid to null geodetic and astronomical observations of space objects in this space-time. We are especially interested in the observed manifestations of anisotropy. The corresponding results are given in the section \ref{s3}. They can be used as tasks for students studying general relativity.

 From the concrete results obtained in the article, we can draw the following general conclusion. For nearby objects, manifestations of the existing anisotropy may go unnoticed against the background of effects caused by the existence of a large-scale structure. But the apparent distribution of distant objects over the sky and over photometric distances makes it possible to detect the existence of anisotropy. First of all, it manifests itself in the anisotropy in the distribution of the CMB flux over the sky and the luminosities and densities of the most distant galaxies and quasars. 
 
 So, the anisotropy manifests itself more clearly for distant objects. In addition, light from the most distant sources, including CMB, was emitted when the growth of inhomogeneities had only just begun and LSS didn't have time to form. Therefore, the influence of the effects caused by LSS does not distort the observed picture. We shown in this article that the anisotropy cannot be weak for the case of the Kasner metric. Therefore, it should manifest itself in observations of both CMB and the most distant objects.

The comparison of these results with the properties of the real Universe known from astronomical observations does not yet reveal manifestations of effects associated with anisotropy, even for sources close to the cosmological horizon, the light from which was emitted during the recombination epoch. This may mean either that our Universe is always isotropic from the Big Bang or that it has become almost isotropic before the recombination epoch, for example, during the inflationary epoch, as described in the article \citep{p3}. The observations do not contradict the last possibility based on the equations of general relativity and on the assumption of the existence of cosmological inflation. 

Should manifestations of space-time anisotropy happen to be discovered in astronomical observations, this would offer a choice of two possibilities. The first is that the inflationary stage did not exist. The second is that when describing this stage, we must replace general relativity with some alternative theory of gravity. Since no manifestations of anisotropy were found in astronomical observations, including the study of the CMB, we are not faced with such a choice. However, it is possible that the problem of anisotropy will become more relevant with progress in the observation of objects in the early stages of the Universe, including the first stars and even Dark Ages. For now it can be assumed that since the end of the inflationary stage, the Universe has always been isotropic or almost isotropic. But this does not exclude the possibility of its significant anisotropy at the moment of birth, during the time of the Big Bang.

{\bf Data availability statement}

No new data were created or analysed in this study.

{\bf Acknowledgement}

The work was partially supported by the National Research Foundation of Ukraine under Project No. 2020.02/0073. I thank Ruth Durrer for valuable discussions and the D\'epartement de Physique Th\'eorique ofthe  Universit\'e de Gen\`eve for their hospitality.

{\bf ORCID ID}

Serge Parnovsky https://orcid.org/0000-0002-1855-1404

\end{document}